\documentclass[aoas,nameyear,dvips]{arximspdf}
\usepackage[dvips,all]{xy}
\usepackage{graphicx}

\doi{10.1214/10-AOAS408}
\volume{5}
\issue{4}
\pubyear{2011}
\firstpage{2266}
\lastpage{2277}

\begin{document}
\begin{frontmatter}

\title{The duality diagram in data analysis: Examples of modern applications}
\runtitle{Introduction to the duality diagram}

\begin{aug}
\author[A]{\fnms{Omar} \snm{De la Cruz}\ead[label=e1]{odlc@stanford.edu}\thanksref{aut1}}
and
\author[A]{\fnms{Susan} \snm{Holmes}\corref{}\ead[label=e3]{susan@stat.stanford.edu}\ead[label=u1,url]{http://www-stat.stanford.edu/\textasciitilde susan/}\thanksref{aut2}}
\runauthor{O. De la Cruz and S. Holmes}
\thankstext{aut1}{Supported in part by NSF Grant DMS-EMSW21-VIGRE-0502385.}
\thankstext{aut2}{Supported in part by NIH Grant R01-GM086884.}
\affiliation{Stanford University}
\address[A]{Department of Statistics\\
Stanford University\\
Sequoia Hall\\
Stanford, California 94305 \\
USA\\
\printead{e1}\\
\hphantom{E-mail: }\printead*{e3}\\
\printead{u1}} 
\end{aug}

\received{\smonth{8} \syear{2010}}

%
\begin{abstract}
Today's data-heavy research environment requires the integration of
different sources of information into structured data sets that can not
be analyzed as simple matrices.
We introduce an old technique, known in the European data analyses circles
as the Duality Diagram Approach, put to new uses through the use of a
variety of
metrics and ways of combining different diagrams together.
This issue of the Annals of Applied Statistics
contains contemporary examples of how this approach provides
solutions to hard problems in data integration. We present here the
genesis of the technique
and how it can be seen as a precursor of the modern kernel based approaches.
\end{abstract}


\vspace*{-5pt}
\end{frontmatter}

\section{Introduction}
Multivariate statistical methods have been used for ma\-ny decades to
deal with situations in which two or more variables are measured
or recorded for each unit.

A classical example of this situation is Guerry's data set, in which
several variables
meant to capture ``moral qualities'' (e.g., literacy, crime
rate, suicide rate)
were tabulated for each of the departments in which France was divided
at the
time (1833). This data set suggests that one can be interested in how the
variables change as one moves around in France, or one can be interested
in how the departments compare to each other based on the measured
characteristics.
It is of special interest how these two approaches can be combined;
this is
considered in detail in \citet{Dray-Jombart2010}.

Besides having a combination of two essentially multidimensional
sources of
information, like geographic location plus recorded data, another layer of
complexity is added when one more variable like time is added, leading to
what essentially are two or more data cubes. A typical example of this
is
ecological data, where species abundances are measured at different, specified
locations,\vadjust{\goodbreak} over the course of time. The different approaches used in
this setting
are reviewed in \citet{Thioulouse2010}, using the duality diagram
setting as a
unifying framework.

Such an approach is not limited to animal or plant species spread over
a~geographic area; the advent of \textit{metagenomics} has made it possible
to study the abundances of bacterial species in locations like ocean or pond
water, or even the human gut. In this case it becomes important to incorporate
information not only about location in space, but also in the
phylogenetic landscape,
by using established or inferred phylogenetic trees for the bacterial species
detected. This problem is addressed, using the duality diagram formalism,
in \citet{Purdom2010}.

As an outgrowth of methods favored by French statisticians,
\citet{CP}
proposed a unifying framework capable of
including many methods reinvented and used by different groups in
different countries
as special cases. This framework is based on the analysis of certain
linear operators
between inner-product spaces which can be naturally associated to a
data matrix, in
the same way Kernel matrices are used today in machine learning
\citet{kernelpca1}.
This is explained in detail in works like \citet{Escoufier2006} and \citet{holmes2006maf}.
In this article we present some motivations behind the choices
made for this approach in the accompanying papers.

The notion of duality is everywhere in Mathematics, appearing under
different guises in
the most diverse fields; and it is often remarkably useful.
The idea of duality was introduced in the analysis of multivariate data
by the French school
of data analysts as a way to unify a suite of methods that turned out
to be
exactly or almost exactly equivalent to methods known by a different name,
and the duality-diagram formalism provides a
simple way to put all these methods in the same context.

Since this approach is the basis of the special articles presented
together here [\citet{Dray-Jombart2010}, \citet{Purdom2010},
 \citet{Thioulouse2010}], this short introduction aims to establish
the basic facts and notation.
The abstract approach in the duality diagram setup is often
intimidating and it possibly
turns away some interested readers; we hope we can show here that these
notions are
actually natural, and that the overhead due in understanding the
notation pays off
handsomely in the breadth and complexity of applications.

\vspace*{-3pt}\section{The data matrix as an operator between inner-product spaces}
Today the distinction between the space of rows of the matrix as a
sample from a population and the space of columns as the fixed
variables on which
the observations were measured has been softened and we often hear the term
`transposable' data. The definitions presented here explain this
row-column duality.

By dispensing of the traditional probabilistic sample-population
interpretation, European data\vadjust{\goodbreak} analysts in the 1970s [\citet{Benzecri1973}, \citet{CP}, \citet{Gifi}] can be seen in hindsight as
precursors of the current Machine learning schools.
It is interesting to remember that all these schools had precursors
who spent time
at the AT\&T laboratories in New Jersey at a time when John Tukey was
active there.

Consider an $n\times p$ matrix $\mathbf{X}$ containing data for
variables $V_1,\dots,V_p$
collected from $n$ individuals or units. This matrix defines an operator
$L_\mathbf{X}\dvtx\mathbb{R}^p \to\mathbb{R}^n$ by the rule $\mathbf
{v}\mapsto\mathbf{B}\mathbf{v}$.
What interpretation can we give to such a map? The vector $\mathbf{v}$
can be considered
to contain the coefficients for linearly\vadjust{\goodbreak} combining the variables
$V_1,\dots,V_p$ into a
new, synthetic, variable. In that sense, it becomes apparent that
actually we should consider
this a map from ${\mathbb{R}^p}^*$, the dual space of $\mathbb{R}^p$,
into $\mathbb{R}^n$.
The map $L_\mathbf{X}$ provides a way to fill in the $n$ values for the
new synthetic variable
$V=v_1V_1+\cdots+v_pV_p$, which could have been defined even before
collecting the data.
From now on we will abuse notation and identify the operator $L_\mathbf
{X}$ and the matrix
$\mathbf{X}$ (and will do the same with other similarly defined
operators and matrices). We
have then the following portion of the diagram:
\[
\xymatrix{
{\mathbb{R}^p}^* \ar[r]^{\mathbf{X}} & \mathbb{R}^n. }
\]
%
\subsection{Adjoint operators as a useful formalism}
Recall that the adjoint of a linear transformation $T\dvtx\mathbb{V}_1\to
\mathbb{V}_2$ between inner
product spaces is defined as the mapping $T^*\dvtx\mathbb{V}_2\to\mathbb
{V}_1$ that satisfies
\[
\langle Tu,z\rangle_2=\langle u,T^*z\rangle_1\qquad  \forall u\in V_1,
\forall z\in V_2
\]
(for simplicity, we will only
consider spaces with scalars in $\mathbb{R}$ in this article; this is
enough for most data analyses).
This can be seen as just a clever way of extending the notion of matrix
transpose to a more
general setting, but it is actually a powerful formalism, especially
when dealing with multiple
inner products on the same spaces (notice that $T^*$ depends not only
on $T$ but also on
$\langle\cdot,\cdot\rangle_1$ and $\langle\cdot,\cdot\rangle_2$).

It is convenient sometimes to think of $T^*$ as a map from $\mathbb
{V}_2^*$ to $\mathbb{V}_1^*$; this matches
the corresponding situation when it is generalized to Banach spaces. In
our setting this
distinction might be considered moot, since all spaces considered are
naturally isomorphic to their
duals, but we will continue using the star notation; the diagrams, and
all the matrix operations
obtained from them, work equally well if the stars are dropped from the spaces.
Then we have the following:
\[
\mbox{
\includegraphics{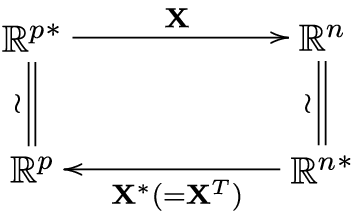}\vadjust{\goodbreak}
}
\]
Since we are considering the standard inner products on $\mathbb{R}^p$
and $\mathbb{R}^n$
(and their dual spaces), $\mathbf{X}^*$ corresponds just to the
transpose $\mathbf{X}^T$ of $\mathbf{X}$.
Thus, $\mathbf{X}^*\mathbf{X}=\mathbf{X}^T\mathbf{X}$, $\mathbf
{XX}^*=\mathbf{XX}^T$,
these two symmetric matrices have the same eigenvalues (except possibly
for zeros to account
for the difference between~$p$ and $n$), and the two sets of
eigenvectors can be used to form the singular value
decomposition (SVD) of $\mathbf{X}$ [\citet{golub-vanloan1996}].

\vspace*{-4pt}\subsection{The general duality diagram}
Consider now the situation where the inner products (i.e., the geometries)
on $\mathbb{R}^p$ and $\mathbb{R}^n$
are not standard. That is, assume that there are symmetric, positive
definite matrices~$\mathbf{Q}_{p\times p}$ and~$\mathbf{D}_{n\times n}$ such that the inner products\vspace*{-1pt}
\[
\langle\mathbf{u},\mathbf{v}\rangle_\mathbf{Q} := \mathbf{u}^T\mathbf
{Q}\mathbf{v}\qquad
\forall\mathbf{u},\mathbf{v}\in\mathbb{R}^p\vspace*{-1pt}
\]
and\vspace*{-1pt}
\[
\langle\mathbf{w},\mathbf{z}\rangle_\mathbf{D} := \mathbf{w}^T\mathbf
{D}\mathbf{z}\qquad
\forall\mathbf{w},\mathbf{z}\in\mathbb{R}^n\vspace*{-1pt}
\]
somehow make more sense for a particular data analysis than the
standard inner products.
A typical example is when $\mathbf{D}$ is a diagonal matrix of
(positive) weights, one for each
individual, down-weighting individuals that are known to have been
measured with a larger error;
another example is when $\mathbf{Q}$ is the diagonal matrix containing
the reciprocals of the
sample variances
for the columns of $\mathbf{X}$, which corresponds to standardizing the
variables (assuming
they are already centered); a~related example is when $\mathbf{Q}$
is the inverse of the sample variance--covariance matrix
obtained from $\mathbf{X}$, in which case the new geometry corresponds
to the \textit{Mahalanobis
distance}. Often, we want to consider the case in which $\mathbf
{Q}=\mathbf{LL}^T$ and
we are interested in a set $\{U_1,\dots,U_p\}$ of transformed variables
obtained from $\{V_1,\dots,V_p\}$
by multiplication by $\mathbf{L}$, leading to a transformed data matrix
$\mathbf{Y}=\mathbf{XL}$.

Different multivariate procedures can be obtained by appropriately
choosing $\mathbf{Q}$
and $\mathbf{D}$; see Section \ref{section:particular-cases} for some examples.

Instead of $\mathbf{X}$ and its adjoint, consider now the transformation
$\mathbf{XQ}\dvtx\mathbb{R}^p\to\mathbb{R}^n$. That is, a vector $\mathbf
{v}$ of coefficients is first
transformed into $\mathbf{L}^T\mathbf{v}$, which is in the scale of the
transformed data matrix
$\mathbf{Y}=\mathbf{XL}$,
and then used to create a linear combination of
the variables $U_1,\dots,U_p$. Then, for all $\mathbf{u}\in\mathbb
{R}^p$, $\mathbf{z}\in\mathbb{R}^n$,\vspace*{-1pt}
\[
\langle\mathbf{XQu},\mathbf{z}\rangle_\mathbf{D} = (\mathbf
{XQu})^T\mathbf{Dz}
= \mathbf{u}^T\mathbf{QX}^T\mathbf{Dz}
=\langle\mathbf{u},\mathbf{X}^T\mathbf{Dz}\rangle_\mathbf{Q},\vspace*{-1pt}
\]
so $(\mathbf{XQ})^*=\mathbf{X}^T\mathbf{D}$. Then, $(\mathbf
{XQ})^*\mathbf{XQ}=\mathbf{X}^T\mathbf{DXQ}$
and
$\mathbf{XQ}(\mathbf{XQ})^*=\mathbf{XQX}^T\mathbf{D}$ are self-adjoint
operators on
$(\mathbb{R}^p,\langle\cdot,\cdot\rangle_\mathbf{Q})$ and
$(\mathbb{R}^n,\langle\cdot,\cdot\rangle_\mathbf{D})$, respectively,
but they are not necessarily symmetric matrices.
Nevertheless, they have real eigenvalues (which match, except for zeros
to account for the difference
between $p$ and $n$), because they are similar to symmetric matrices by
way of positive definite
matrices; for example,\vspace*{-1pt}
\[
\mathbf{Q}^{1/2}(\mathbf{X}^T\mathbf{DXQ})\mathbf{Q}^{-1/2}=\mathbf
{Q}^{1/2}\mathbf{X}^T\mathbf{DX}\mathbf{Q}^{1/2},\vspace*{-1pt}
\]
where $\mathbf{Q}^a$ is obtained by replacing each eigenvalue\vadjust{\goodbreak} $\lambda$
of $\mathbf{Q}$ with $\lambda^a$.

The eigenvectors of $\mathbf{X}^T\mathbf{DXQ}$ and $\mathbf
{XQX}^T\mathbf{D}$ are also real; however, they
need not be orthogonal. Nevertheless, the eigenvectors of $\mathbf
{X}^T\mathbf{DXQ}$ can be taken
to be orthogonal with respect to $\langle\cdot,\cdot\rangle_\mathbf
{Q}$, and those of $\mathbf{XQX}^T\mathbf{D}$
to be orthogonal with respect to $\langle\cdot,\cdot\rangle_\mathbf
{D}$. (This can be interpreted as
leading to a~generalized version of the SVD.)

In diagram form, we have the following:
\[
\mbox{
\includegraphics{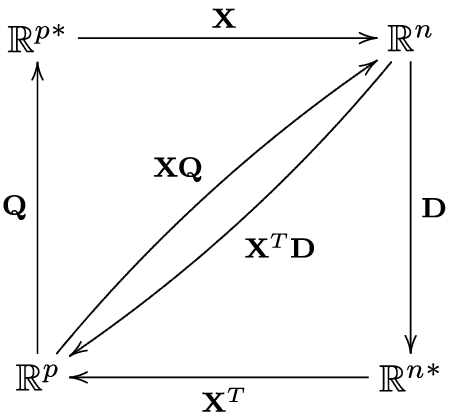}
}
\]
%

This way, the triplet of matrices $(\mathbf{X},\mathbf{Q},\mathbf{D})$
defines a multivariate
data analysis setup, in which the main strategy is the computation of
the eigendecompositions of
the matrices $\mathbf{X}^T\mathbf{DXQ}$ and $\mathbf{XQX}^T\mathbf{D}$.
It is also customary to denote $\mathbf{V}=\mathbf{X}^T\mathbf{DX}$ and
$\mathbf{W}=\mathbf{XQX}^T$,
so that the two operators of interest become $\mathbf{VQ}$ and $\mathbf{WD}$.

The eigendecomposition
can be computed for the smaller of the two matrices (which is usually
much smaller than the other), and, if needed,
the eigenvectors for the other one can be easily obtained: for example, if
$\mathbf{X}^T\mathbf{DXQv}=\lambda\mathbf{v}$, then $\mathbf{w}:=\mathbf
{XQv}$ satisfies
\[
\mathbf{XQX}^T\mathbf{Dw}=\mathbf{XQ}\bigl( \mathbf{X}^T\mathbf{DXQv}
= \mathbf{XQ}(\lambda\mathbf{v}) \bigr)= \lambda\mathbf{w}.
\]
Furthermore, orthogonality is also preserved among eigenvectors: if
$\mathbf{v}_1,\mathbf{v}_2$ are
$\mathbf{Q}$-orthogonal eigenvectors for $\mathbf{X}^T\mathbf{DXQ}$, then
\[
\langle\mathbf{XQv}_1,\mathbf{XQv}_2\rangle_\mathbf{D} =
\mathbf{v}_1^T\mathbf{QX}^T\mathbf{DXQv}_2
= \mathbf{v}_1^T\mathbf{Q}(\lambda\mathbf{v}_2)
=\lambda\langle\mathbf{v}_1,\mathbf{v}_2\rangle_\mathbf{Q}
=0.
\]
Thus, whole eigendecompositions are easily transferred.

\subsection{Connections with kernel methods}
The operator $\mathbf{XQX}^T\mathbf{D}$ can be seen as a precursor of
the more general kernel
matrices used in today's kernel PCA type methods [\citet{kernelpca1}].

In the kernel approach to data analysis one assumes that the data is provided
as a $n\times n$ matrix $\mathbf{K}$
containing
proximity scores for each pair of individuals; these scores might have
been computed
from measured variables (as in the case of the matrix $\mathbf
{XQX}^T\mathbf{D}$,
the \textit{linear kernel}; nonlinear functions of the variables offer a
great variety of
other possibilities), or by directly
comparing the individuals; see \citet{Scholkopf2004kx} for a review of
the theory
and examples of applications in computational biology.\vadjust{\goodbreak}

Kernel matrices are similarity matrices, that is, the proximity score
for two individuals
is high when they are similar and low (even negative) when they are
dissimilar. (Distance
matrices, on the other hand, have higher values for dissimilar pairs of
individuals.)
What is the meaning of $\mathbf{K}$ as an operator on $\mathbb{R}^n$?
A useful interpretation is to consider it as a smoothing operator
acting on real-valued functions $f$
defined on the set of individuals: the value of $\mathbf{K}f$ at the
$i$th individual
is a weighted sum of all the values of $f$, with higher weights
for those individuals more similar to the $i$th; in other words, $f$ is averaged
locally (up to a multiplicative constant). This offers one explanation
of why the
eigendecomposition of $\mathbf{K}$ is useful, since repeated
application of $\mathbf{K}$
to $f\in\mathbb{R}^n$
(which should produce a very smooth function) converges toward an eigenvector
of the leading eigenvalue, and ``very smooth'' functions of the data
points can be used
as coordinates.

\citet{Purdom2010} explores the similarities between the
duality diagram and kernel approaches in
Appendix B, for the case of kernel Canonical Correspondence Analysis.

\section{Examples of well-known methods as particular cases of the
diagram}\label{section:particular-cases}
Here we briefly describe how some well-known multivariate methods can
be expressed as
particular cases of the duality diagram, by appropriately choosing
$\mathbf{Q}$ and $\mathbf{D}$.
We will assume that $\mathbf{X}$ is centered by columns (i.e., the
mean has been subtracted for each variable).

\subsection{Principal components analysis (PCA)}
PCA seeks to find linear combinations of the variables that explain
most of the variability in the
data; see \citet{Mardia1979ri}, for example, for more details.

Take $\mathbf{Q}=\mathbf{I}_p$, and $\mathbf{D}=\frac{1}{n}\mathbf
{I}_n$. This corresponds
to PCA in the original scales; it is equivalent to a straightforward
SVD on $\mathbf{X}$ (except for
the factor~$1/n$).

If one standardizes the variables, as it is often appropriate to
eliminate unit scale effects, then
$\mathbf{Q}$ is taken to be the diagonal matrix containing the
reciprocals of the sample
variances of the columns of $\mathbf{X}$
(so $\mathbf{L}$ contains the reciprocals of the standard deviations).
While the $i$th eigenvector $\mathbf{v}_i$
of the ($\mathbf{D}$-weighted) sample covariance matrix $\mathbf
{Y}^T\mathbf{DY}$ provides the
loadings of the variables $U_1,\dots,U_p$ for the $i$th principal
component (so that the actual components have
to be obtained by $\mathbf{p}_i=\mathbf{Yv}_i/\sqrt{\lambda_i}$),
$\mathbf{p}_i$ can be obtained directly as
an eigenvector for $\mathbf{XQX}^T\mathbf{D}$: indeed,
\begin{eqnarray*}
\mathbf{XQX}^T\mathbf{Dp}_i &=& \mathbf{XQX}^T\mathbf{DYv}_i /\sqrt
{\lambda_i}
= \mathbf{XLL}^T\mathbf{X}^T\mathbf{DYv}_i/\sqrt{\lambda_i}\\
&=& \mathbf{YY}^T\mathbf{DYv}_i /\sqrt{\lambda_i}
= \mathbf{Y}\lambda_i\mathbf{v}_i /\sqrt{\lambda_i}
= \lambda_i \mathbf{p}_i.
\end{eqnarray*}
Computing the principal components $\mathbf{p}_i$ does not require the
explicit decomposition of $\mathbf{Q}$
as $\mathbf{LL}^T$.
\subsection{Correspondence analysis (CA)}
A total of $m$ observations are classified according to two categorical
variables, one with $n$ categories or
levels, and the other with $p$, producing a $n\times p$ matrix $\mathbf
{N}$ of counts for each combination of
levels (a \textit{contingency table}). One wants to study how the counts
differ from the expected counts
under the assumption of independence between the two variables. To cast
CA as a duality diagram, we first
define the frequency matrix $\mathbf{F}=\mathbf{N}/m$ and the marginal
frequency vectors
$\mathbf{c}=\mathbf{F}^T\mathbf{1}_{n\times1}$ and
$\mathbf{r}=\mathbf{F}\mathbf{1}_{p\times1}$; then, the expected
counts (conditionally on the marginals)
are given by $n\mathbf{rc}^T$. Using the matrices $\mathbf{D}_r=\textrm
{diag}(\mathbf{r})$ and
$\mathbf{D}_c=\operatorname{diag}(\mathbf{c})$, we can standardize $F$ by
\[
\mathbf{X}:= \mathbf{D}_r^{-1}(\mathbf{F}-\mathbf{rc}^T)\mathbf{D}_c^{-1}=
\mathbf{D}_r^{-1}\mathbf{F}\mathbf{D}_c^{-1}-\mathbf{1}_{n\times p}.
\]
The matrix $\mathbf{X}$ seems like a reasonable choice to study by
eigendecomposition. However, all
rows and columns have been reduced to the same importance, while,
heuristically, categories with larger marginal
counts should provide more accurate information on the distribution of
the other variable, and thus should be
given greater weight. This can be achieved by defining the triplet
$(\mathbf{X},\mathbf{D}_c,\mathbf{D}_r)$.
Notice that actually $\mathbf{X}$ is centered by rows and by columns
with respect to the inner products
given by $\mathbf{D}_r$ and $\mathbf{D}_c$. This approach matches the
traditional definition of CA.
\citet{Purdom2010} shows how the information about relationships between
the rows
of the contingency table can be incorporated into the duality diagram
in the special
case where there are binary trees that connects the rows of the
abundance matrix.

\subsection{Variance, inertia, co-inertia}
The study of variability of one continuous variable is done through the
use of the variance;
this notion is generalized in several different directions
to accommodate the complexities of dealing with multiple tables,
graphs, etc.,
through the concept of inertia.
As in physics, we define inertia as a weighted sum of squared distances
of the weighted points.
For each of the diagrams studied above, the inertia designates the
trace of the
operator $WD$, and we have $\mbox{Inertia}_{\mathrm{total}}=\operatorname{tr}(WD)=\operatorname{tr}(VQ)$. As pointed out in \citet{Purdom2010}, in the
case of CA, the inertia is proportional to the $\chi^2$ statistic,
whereas in ordinary PCA it is just the total variance of all the
variables. In discriminant analysis, the inertia is decomposed into
between-groups and within-group components; these are also used in the
BCA analysis [\citet{Thioulouse2010}, \citet{Dray-Jombart2010}].

The weighted distances between columns have another interpretation in
ecology and
\citet{Purdom2010} shows how they can be associated to different
measures of diversity.

The decomposition of total inertia can be seen as a generalization to
MANOVA which is
the special case of a variance decomposition. \citet{Purdom2010} uses
this effectively to
show how to decompose the total diversity across all locations
into the average diversity of individual locations
and plus the average of pairwise dissimilarities of locations.

\citet{Dray-Jombart2010} use similar decompositions to show what
part of the inertia can be assigned to spatially local variation in
their BCA approach to multivariate spatial data.
They also show how the graphical relationships between rows can be
encoded in a
special metric $\mathbf{D}$ built from the weighted connectivity matrix.
(In their paper, they call these weights $W$.)


\section{One more level of complexity: Comparing diagrams}
Interesting results can be obtained by combining two or more triplets.
The usual assumption is that
two (or more) sets of variables are measured on the same set of~$n$
individuals; thus, the matrix $\mathbf{D}$
is assumed to be common, but each set of variables has its own version
of $\mathbf{Q}$, of the appropriate
size.

For example, one of the triplets might contain data from variables
measured on each of the individuals,
while the other might encode known relationships between the individuals.

\subsection{The RV coefficient}
A key element in the comparison of the operators arising from two
duality diagrams is the RV coefficient.
It can be considered as a generalization of the squared correlation coefficient
by using the Froebenius matrix product.

Given two symmetric matrices $\mathbf{A},\mathbf{B}$ of the same size,
we define
$\operatorname{COVV}(\mathbf{A},\allowbreak\mathbf{B})=\operatorname{tr}(\mathbf{AB})$, and
\[
\operatorname{RV}(\mathbf{A},\mathbf{B})=\frac{\operatorname{tr}(\mathbf{AB})}{\sqrt
{ \operatorname{tr}(\mathbf{AA})\operatorname{tr}(\mathbf{BB}) }},
\]
whenever $\mathbf{A},\mathbf{B}\neq\mathbf{0}$.
Many nice properties of these definitions arise from the fact that
$\operatorname{tr}(\mathbf{AB})$
defines an inner product on the vector space of symmetric matrices of a
given size.
This can be adapted to the general setting of multiple duality
diagrams: having $\mathbf{D}$ fixed,
call $S(\mathbf{D})$ the vector space of $\mathbf{D}$-symmetric
matrices, that is, matrices
satisfying $\mathbf{DA}=\mathbf{A}^T\mathbf{D}$ (equivalently,~$\mathbf
{A}$ is self-adjoint
with respect to $\langle\cdot,\cdot\rangle_{\mathbf{D}}$). Then $\operatorname{tr}(\mathbf{AB})$
defines an inner product on $S(\mathbf{D})$.

When comparing two duality diagrams
$(\mathbf{X_1},\mathbf{Q}_1,\mathbf{D}),(\mathbf{X_2},\mathbf
{Q}_2,\mathbf{D})$, then numbers
$p_1,p_2$ of variables might be different, yielding matrices $\mathbf
{V}_1\mathbf{Q}_1, \mathbf{V}_2\mathbf{Q}_2$
of different size; however, we will be comparing the matrices
(operators)~$\mathbf{W}_1\mathbf{D}$ and
$\mathbf{W}_2\mathbf{D}$, which are of the same size and $\mathbf{D}$-symmetric.
We define the RV coefficient of the two diagrams as $\operatorname{RV}(\mathbf
{W}_1\mathbf{D},\mathbf{W}_2\mathbf{D})$.

Some immediate properties of the RV coefficient are as follows: its
values are always in $[0,1]$; it equals 1
only when $\mathbf{W}_1=\alpha\mathbf{W}_2\neq\mathbf{0}$, for some
nonzero scalar~$\alpha$; and it equals 0 only when
$\mathbf{X}_1^T\mathbf{DX}_2=\mathbf{0}$ (provided $\mathbf{Q}_1,\mathbf
{Q}_2$ are nonsingular).
The proofs are not too hard; more details can be found in \citet{Escoufier2006}.

The RV coefficient between diagrams (or triplets) can be used for
justifying the use of eigenvalues
and eigenvectors in this setting. For example, performing PCA based on
$(\mathbf{X},\mathbf{Q},\mathbf{D})$
and selecting the $q$ top components is equivalent to finding a matrix
$\mathbf{Z}_{n\times q}$ such that
the RV coefficient between $(\mathbf{X},\mathbf{Q},\mathbf{D})$ and
$(\mathbf{Z},\mathbf{I}_q,\mathbf{D})$
is maximized.

\subsubsection{PCA with respect to instrumental variables}
When one data set~$\mathbf{Y}$ has the special status of a response
that we would like to predict or explain
from the other data set $\mathbf{X}$ of explanatory variables, we can
generalize ordinary regression to a multivariate response through the
same diagram framework.
This is called \textit{PCA with respect to instrumental variables},
abbreviated PCA-IV (also known as redundancy analysis, RDA), first
described by \citet{Rao1964}.
In terms of the comparison of duality diagrams and RV coefficients,
this problem can be rephrased as that of
finding the metric $\mathbf{M}$ to associate to $\mathbf{X}$ so that
$(\mathbf{X}, \mathbf{M}, \mathbf{D})$
is {\em as close as possible} to $(\mathbf{Y}, \mathbf{Q}, \mathbf
{D})$ in the RV sense. That is, we want to maximize
$RV(\mathbf{XMX}^T\mathbf{D},\mathbf{YQY}^T\mathbf{D})$. We abbreviate
the cross-products by writing
\[
\mathbf{X}^T \mathbf{DX} = \mathbf{S}_{xx},\qquad
\mathbf{Y}^T\mathbf{DY}  = \mathbf{S}_{yy},\qquad
\mathbf{X}^T\mathbf{ DY}= \mathbf{S}_{xy}
\]
and
\[
\mathbf{R}  = \mathbf{S}^{-1}_{xx} \mathbf{S}_{xy} \mathbf {Q S}_{yx} \mathbf{S}_{xx}^{-1} .
\]
Then for any $\mathbf{R}$
\[
\Vert\mathbf{YQY}^T\mathbf{D}\!-\!\mathbf{X MX}^T\mathbf{D} \Vert
^2\!=\!\Vert\mathbf{YQY}^T\mathbf{D}\!-\!\mathbf{XRX}^T\mathbf{D}\Vert
^2\!+\!\Vert\mathbf{XRX}^T\mathbf{D}\!-\!\mathbf{XMX}^T\mathbf{D} \Vert^2.
\]
The first term on the right-hand side does not depend on $\mathbf{M}$,
and the second term will be zero
for the choice $\mathbf{M}=\mathbf{R}$.

If we add the extra constraint that we only allow ourselves a rank $q$
approximation, with
$q < \min\{\operatorname{rank} (\mathbf{X}), \operatorname{rank} (\mathbf
{Y})\}$, the optimal choice of a~positive definite matrix $\mathbf{M}$ is to take $\mathbf{M}= \mathbf
{RBB}^T\mathbf{R}$ where
the columns of~$\mathbf{B}$ are the eigenvectors of $\mathbf
{X}^T\mathbf{DXR}$ with
\begin{eqnarray}
\mathbf{B} = \biggl(\frac{1}{\sqrt{\lambda_1}}\bolds{\beta}_1, \ldots,
\frac{1}{\sqrt{\lambda_q}} \bolds{\beta}_q \biggr)\nonumber \\
\eqntext{\mbox{ such that }
\cases{
\mathbf{ X}^T \mathbf{DXR} \bolds{\beta}_k = \lambda_k \boldsymbol
{\beta}_k,\cr
\bolds{\beta}^T_k \mathbf{R}\bolds{\beta}_k = \lambda_k,&\quad
$k=1,\ldots, q$,\cr
\lambda_1 > \lambda_2 > \cdots> \lambda_q.
}}
\end{eqnarray}
The
PCA with regards to instrumental variables of rank $q$ is equivalent
to the PCA of rank $q$ of the triple $(\mathbf{X}, \mathbf{R}, \mathbf
{D}) $ where
\[
\mathbf{R} = \mathbf{S}_{xx}^{-1} \mathbf{S}_{xy} \mathbf{QS}_{yx}
\mathbf{S}_{xx}^{-1}.
\]

\subsection{Comparing more than two diagrams}
Consider $k$ diagrams $(\mathbf{X_1},\mathbf{Q}_1,\allowbreak\mathbf{D}),\dots
,(\mathbf{X_k},\mathbf{Q}_k,\mathbf{D})$.
This could correspond, for example, to $k$ different studies on the
same subjects, using different variables,
or the same variables measured
on the same units at different points in time (time course study); a
review of this problem in the setting of
community ecology is found in \citet{Thioulouse2010}. It is often
important to summarize
the relationships between the diagrams in a compact and intelligible way.
The RV coefficients in fact allow us to consider this as performing a
PCA of the PCAs.
We compute the multivariate correlation coefficients between tables and
use those as the matrix to
be diagonalized, similarly to what happens in ordinary PCA.

The values of the pairwise computations of the COVV and RV coefficients
are arranged into
$k\times k$ symmetric matrices $\mathbf{C}$ and $\mathbf{R}$,
respectively, and the eigendecomposition
of these matrices can lead to useful low-dimensional representations,
just as in the case of PCA using
the covariance or correlation matrices, respectively. In this case, a
2- or 3-dimensional plot can be created in which
each point represents one of the studies (diagrams).

Furthermore, since $\mathbf{C}$ and $\mathbf{R}$ have nonnegative
entries, the eigenvector $\mathbf{u}_1$
corresponding
to the largest eigenvalue can be taken to have only nonnegative
entries, adding up to 1.
Then, defining $\mathbf{W}=\sum_{i=1}^k u_{i1}\mathbf{W}_i$, the
operator $\mathbf{WD}$ can
be taken as a compromise or summary of all the diagrams, and one can
study how far, in the RV sense,
different studies are from the compromise.

These steps are part of the so-called STATIS procedure [\citet{Escoufier1980}].
One can think of these data sets as a data cube,
with three indices; then a~similar procedure can be used to compare two
or more such cubes.

\section{Conclusions}
The duality diagram is a useful formalism that allows one to easily
compare many
classical multivariate methods, revealing what they have in common and where
they differ. But, furthermore, it has become a valuable tool for
dealing with two
problems that have become very common: (1) combining and amalgamating
data which,
although collected from different sources and using different methods,
shed light
on different aspects of the same phenomenon; and (2) taking advantage
of complex,
nontraditional data types, like tree and network information. These
two problems are
closely related, as the data to be amalgamated are often of complex type.

The overhead in effort to understand the abstract definitions in the
duality diagram
approach to data analysis is amply offset by the clearer picture that
is gained and
by the wealth of applications that become available.
In this article we have tried to reduce that overhead by laying out
arguments that
show that those definitions are actually quite natural.
The three articles [\citet{Dray-Jombart2010}, \citet{Thioulouse2010}, \citet{Purdom2010}]
in this group
are excellent examples of the power of this approach, but are only a
small sample
from a large and growing body of work.

Recently,
\citet{shinkareva2008using}
have used the RV coefficient and STATIS approaches to explore
fMRI brain activation in conjunction with stimulations such as images
of tools.

A series of papers [\citet{culhane2002between}, \citet{culhane2003cross}, \citet{fagan2007multivariate}]
have applied BCA and Co-Inertia analyses to the problem of integrating
multiple sources of data from heterogeneous gene expression and
proteomic studies.

\citet{baty2006analysis} used the PCAIV method to identify special genes
in microarray data and
\citet{baty2008stability} used bootstrap and permutation type tests for
evaluating the stability of the gene identifications produced.

Most of the methods presented in these papers have been coded into functions
for the statistical computation environment R [\citet{r2008}],
many available in the library \texttt{ade4}, for which exemplary
presentations have been
published [see \citet{ade4}, \citet{Drayade4}, \citet{dray2007ade4}]. In the case of \citet{Thioulouse2010}, you can even run
in an interactive way
through all the commands generating each and every plot through the
reproducible website
at \url{http://pbil.univ-lyon1.fr/SAOASOPET/}.


\printaddresses

\end{document}